\newcommand{\ber}{\begin{eqnarray}}
\newcommand{\eer}{\end{eqnarray}}
\newcommand{\re}[1] {(\ref{#1})}

\def\nn{\nonumber}
\def\one{1\!\!\!1}
\def\+{{+\!\!\!+}} 
\def\pp{\mbox{\tiny${}_{\stackrel\+ =}$}} 
\newcommand{\half}{\textstyle\frac 1 2}
\newcommand{\ihalf}{\textstyle\frac \imath 2}

\documentclass{PoS}
\usepackage{yfonts}

\title{Semichirals and Four Dimensional Geometry}

\ShortTitle{Semichirals and Four Dimensional Geometry}

\author{\speaker{Ulf Lindstr\"om}
       \thanks
        {Preprint numbers: UUITP-09/15, Imperial-TP-UL-2015-01}\\
       Uppsala University, Sweden and Imperial College, UK.\\
       E-mail: \email{ulf.lindstrom@physics.uu.se}}

\abstract{Semichiral $(2,2)$  sigma models with $4d$ target space are discussed. A novel description in $(1,1)$ superspace allows an analysis of possible extended supersymmetries. It is argued that a manifest semichiral realization of an extra supersymmetry is only possible for hyperk\"ahler target geometry.  A semichiral formulation of the $SU(2)\otimes U(1)$ WZW model is seemingly a counterexample to this. After deriving the extra supersymmetries of this model in $(1,1)$ superspace it is shown  that they cannot be lifted to transformations of semichirals in $(2,2)$ superspace.}

\FullConference{Proceedings of the Corfu Summer Institute 2014 \\
		 3-21 September 2014\\
		 Corfu, Greece}

\begin{document}

\section{Introduction}
A problem discussed in  a recent paper \cite{Lindstrom:2014bra} is reviewed and expounded on here: Under quite general conditions, a realization of additional supersymmetries on the $(2,2)$ fields of a semichiral sigma model with $4d$ target space  requires hyperk\"ahler target space. Nevertheless the $4d$ WZW model representing $SU(2)\otimes U(1)$ has $(4,4)$ supersymmetry in a (twisted) chiral formulation and this formulation has a semichiral dual with non-trivial torsion.

In this paper we argue that the resolution of this seeming contradiction is that the $(4,4)$ is indeed present in the $(1,1)$ superspace formulation of the semichiral model, but it is incompatible with the chirality coditions of the semichiral fields and hence cannot be lifted to a $(2,2)$ semichiral formulation. This is shown using a novel $(1,1)$ description where the auxiliary spinors are kept and the additional supersymmetries act on these as well as on the coordinate fields.

The plan of the paper is as follows: In section 2. the necessary background concerning superspace and complex geometry is given. Section 3.  states the problem with a realization of extra supersymmetries on semichiral fields. In section 4. duality of the $SU(2)\otimes U(1)$ WZW model is presented as well as an analysis of how the Legendre transform relates the condition for $(4,4)$ in a $4d$ (twisted) chiral sigma model with certain isometries to the hyperk\"ahler condition of its semichiral dual. In particular we find a striking formulation of the latter in the natural dual coordinates. The subsection is not directly relevant for the main line of argument, but is included for the sake of completeness.
Section 5. contains the derivation of the novel formulation keeping the auxiliary spinors, examples of transformations satisfying the ``lifting'' condition as well as the derivation of the extra supersymmetries for the semichiral $SU(2)\otimes U(1)$ WZW model and the final negative result on their lifting.

\section{Background}
\subsection{$(2,2)$ superspace}
\label{two}
Generalized K\"ahler Geometry (GKG), \cite{Gualtieri:2003dx}, is the target space geometry of a two-dimensional non-linear sigma model with a generalized
K\"ahler potential, \cite{Lindstrom:2005zr}, $K$ that is a function of chiral, twisted chiral, left and right semichiral, \cite{Buscher:1987uw}, $(2,2)$ superfields :
\ber
S=\int d^2\xi d^2\theta d^2\bar\theta K(\phi,\chi,\ell,r,...)
\eer
where the dots indicate the complex conjugate fields and $(\xi^\mu, \theta^\pm, \bar\theta^\pm)$ coordinatize $(2,2)$ superspace.
The chirality conditions obeyed by the fields are:
\ber
\bar{\mathbb{D}}_\pm \phi=0~,~~~\bar{\mathbb{D}}_+\chi={\mathbb{D}}_- \chi=0~,~~\bar{\mathbb{D}}_+ \ell=0~,~~
\bar{\mathbb{D}}_-r=0~,
\eer
and the complex conjugate relations. The $(2,2)$ algebra is 
\ber
\{\mathbb{D}_\pm, \bar{\mathbb{D}}_\pm \}=\imath\partial\pp
\eer

To describe a sigma model, the number of left and right semichiral fields must be equal. We shall be interested in the minimal symplectic case; one left and one right field. 
\ber
K=K(\ell, \bar \ell, r , \bar r)=: K(L,R)~.
\eer
The corresponding target space is four (real) dimensional.

\subsection{$(1,1)$ superspace}

The reduction to $(1,1)$ superspace reveals the geometry of the model and is achieved using
\ber
\mathbb{D}_\pm=\half(D_\pm-\imath Q_\pm)~,~~~~~~\bar{\mathbb{D}}_\pm=\half(D_\pm+\imath Q_\pm)~,
\eer
where $D_\pm$ are the $(1,1)$ spinor derivative and $Q_\pm$ the nonmanifest second supersymmetry generators.
The action is reduced as
\ber\label{oneact}
S\to\int d^2\xi D^2Q^2 K_|=\int d^2\xi d^2\theta (Q^2 K)_|
\eer
where $Q^2:=Q_+Q_-$ et.c., a horisontal bar denotes setting the second $\theta$ to zero, and the $(1,1)$ components are defined as
\ber\label{onecomps}
\ell_|=\ell~,~~~Q_+\ell_|= \imath D_+ \ell~, ~~~Q_-\ell_|=:\psi_-~,~~~r_|=r~,~~~Q_+ r_|=:\psi_+~,~~~Q_-r_|=i D_-r
\eer
and their complex conjugate. The fields $\psi_\pm$ are auxiliary $(1,1)$ superfields and have to be integrated out to display the $(1,1)$ sigma model:
\ber
S=\int d^2\xi d^2\theta D_+X^i(g_{ij}+B_{ij})D_-X^j=:\int d^2\xi d^2\theta D_+X^iE_{ij}D_-X^j~,
\eer
where $X^ i$ collectively denote $(\ell,\bar\ell,r,\bar r)$, $g$ is a metric and $B$ an antisymmetric tensorfield. 

The $(1,1)$ algebra is simply
\ber
D^2_+=\imath \partial_\+~,~~~~~D^2_-=\imath \partial_=
\eer
\subsection{Generalized K\"ahler geometry}

The bi-hermitean geometry of  \cite{Gates:1984nk} was reformulated as GKG in \cite{Gualtieri:2003dx}. Briefly, it is characterized by a manifold $M$ carrying two complex structures $J_{(+)}$ and $J_{(-)}$
\ber J_{(\pm)}^2=-\one~,
\eer
and a metric $g$ which is hermitean with respect to both of these
\ber
J_{(\pm)}^tgJ_{(\pm)}=g~.
\eer
Further, the complex structures are covariantly constant with respect to $\Gamma^{(+)}$ and $\Gamma^{(-)}$ respective, two metric connections with torsion:
\ber\nonumber\label{covd}
&&\nabla^{(\pm)}J_{(\pm)}=0\\[1mm]
&&\Gamma^{(\pm)}=\Gamma^0\pm\half g^{-1}H~,\quad H=dB~.
\eer
Here $\Gamma^0$ is the usual Levi-Civita connection and $B$ is identified with the $B$-field accomanying the metric in $E$. The geometric data can thus be taken to be the set 
$(M,J_{(\pm)}, g, H)$, but there are several other ways of characterizing GKG; see  \cite{Lindstrom:2012ns} for a discussion of this in relation to sigma models in different superspaces.

There are certain features of the semichiral (symplectic) description of GKG that we shall need. Defining the matrix of derivatives of $K$ to be $K_{LR}$:
\ber
K_{LR}:=\left(\begin{array}{cc}K_{\ell r}&K_{\ell\bar r}\\
K_{\bar\ell r} & K_{\bar \ell\bar r}\end {array}\right)
\eer
we find the symplectic form
\ber
\Omega:=
\left(\begin{array}{cc}0&K_{LR}\\
-K_{RL}&0\end {array}\right)~.
\eer
Using this and the complex structures, we recover the metric and $B$-field according to
\ber
g=\Omega [J_{(+)},J_{(-)}]~,~~~~B=\Omega \{J_{(+)},J_{(-)}\}~.
\eer
Since $\Omega$ is symplectic, we see that $H=dB=0$ when 
\ber\label{Anticom}
\{J_{(+)},J_{(-)}\}=2c\one~,
\eer
with $c$ constant. (The factor two is inserted for convenience.) Under this condition the torsion vanishes and the geometry is (pseudo)hyperk\"ahler \cite{Lindstrom:2005zr}. This can be explicitly verified from the following set of $SU(2)$ worth of  (pseudo-) complex structures $J^{(1)},J^{(2)},J^{(3)}$, 
\ber\nn\label{hks}
J^{(1)}&:=&\frac {1}{\sqrt{1-c^2}}\left(J_{(-)}+ cJ_{(+)} \right)~,\nn\\
J^{(2)}&:=&\frac 1 {2\sqrt{1-c^2}}[J_{(+)} , J_{(-)} ]~,\\[2mm]
J^{(3)}&:=&J_{(+)} ~.
\eer
For $|c|<1$ the geometry is hyperk\"ahler, while for $|c|>1$ the geometry is  pseudo-hyperk\"ahler  \cite{Goteman:2009xb}.

\section{Additional supersymmetry}
\label{Three}

The question of under which conditions a semichiral sigma model can have extra supersymmetries has been discussed for general targetspace dimensions in \cite{Goteman:2009ye}, and for $4d$ target space in \cite{Goteman:2012qk}. In the latter paper the following ansatz for the extra supersymmetry is made:
\ber\nn\label{ans}
\delta\ell &=&\bar\epsilon^+ \bar{\mathbb{D}}_+f({L}, {R}) + g({\ell})\bar\epsilon^{-}\bar{\mathbb{D}}_{-} {\ell}\, + h({\ell})\epsilon ^- \mathbb{D}_{-}{\ell}~,\\[1mm]\nn
\delta\bar\ell&=&\epsilon^+ {\mathbb{D}}_+\bar f({L}, {R})+\bar g(\bar\ell)\epsilon^-\mathbb{D}_-\bar\ell +\bar h(\bar\ell)\bar\epsilon^-\bar{\mathbb{D}}_-\bar\ell~,\\[1mm]\nn
\delta r &=&\bar\epsilon^- \bar{\mathbb{D}}_-\tilde f({L}, {R}) + \tilde g(r )\bar\epsilon^{+}\bar{\mathbb{D}}_{+} {r}\, + \tilde h({r})\epsilon ^+\mathbb{D}_{+}{r}~,\\[1mm]
\delta \bar r &=&\epsilon^- {\mathbb{D}}_-\bar{\tilde f}({L}, {R}) + \bar{\tilde g}(\bar r )\epsilon^{+}{\mathbb{D}}_{+} {\bar r} + \bar{\tilde h}({\bar r})\bar \epsilon ^+\bar{\mathbb{D}}_{+}{\bar r}~.
\eer

This ansatz does not include central charge transformations which have been seen to be important in similar context, e.g., in \cite{Hull:1985pq}. However, the transformations in \re{ans} can only give an on-shell algebra, so central charges will be irrelevant. To see that one is forced on-shell, note that supersymmetry requires
\ber
[\delta^{(+)}_1, \delta^{(+)}_2]{\ell}=i\epsilon^{+}_{[2}\bar\epsilon^{+}_{1]}\partial{}{\ell},
\eer
 whereas \re{ans} gives
\ber\nn
&&[\delta_1, \delta_2]{\ell}=\\[1mm]\nn
&&-\epsilon^+_{[2}\bar\epsilon^+_{1]}\left(|f_{\bar\ell} |^2 i\partial_{\+}{\ell}+(f_{\bar \ell}\bar f_r+f_r\tilde h) \bar{\mathbb{D}}{+} \mathbb{D}_+r+...\right)\nn\\
&& + \bar\epsilon^-_{[2}\epsilon^-_{1]}(-{gh}) i\partial_= {\ell}+\dots~.
\eer
Since $|f_{\bar\ell} |^2>0$ we will need relations between the fields, i.e.  the field equations.

A further result of \cite{Goteman:2012qk} is that $gh=-1$, which is shown to imply that \re{Anticom} is satisfied with $c$ a constant, i.e., that the geometry is hyperk\"ahler. Although not conclusive, the evidence strongly suggests that a manifest additional supersymmetry of a semichiral model requires a hyperk\"ahler target space. The present work is motivated by the existence of a model that seemingly constitutes a counterexample to this conclusion; the $SU(2)\otimes U(1)$ WZW model in semichiral coordinates.

\section{Duality}
\subsection{The $SU(2)\otimes U(1)$ WZW model in semichiral coordinates}

There is a generalized K\"ahler potential describing the $SU(2)\otimes U(1)$ WZW model that involves chiral and twisted chiral superfields. It reads \cite{Rocek:1991vk}
\ber\label{Bilp1}
K=-ln\hat\chi ln\hat{\bar\chi}=\int^{\frac{\hat\phi\hat{\bar\phi}}{\hat\chi\hat{\bar\chi}}}dq\frac{ln(1+q)}q~,
\eer
and satisfies Laplace's equation
\ber\label{Lap}
K_{\hat\phi\hat{\bar\phi}}+K_{\hat\chi\hat{\bar\chi}}=0~,
\eer
which is the necessary condition for $(4,4)$ supersymmetry \cite{Gates:1984nk}. In addition there is a simple  transformation to new chiral and twisted chiral coordinates, $\phi=ln\hat\phi$ and  $\chi=ln\hat\chi$, where the model is easily dualized:
\ber\label{Bilp2}
K\to K=\half(\chi-\bar\chi)^2+\alpha(\chi-\bar\chi)(\phi-\bar\phi)+\int^{\phi+\bar\phi-\chi-\bar\chi}dq~{ln(1+e^q)}~.
\eer
In the new coordinates, the condition for $(4,4)$ supersymmetry becomes
\ber\label{Lap2}
e^{-(\phi+\bar\phi)}K_{\phi {\bar\phi}}+e^{-(\chi+\bar\chi)}K_{\chi {\bar\chi}}=0~.
\eer
To facilitate dualization, the $K$ in \re{Bilp2}  is the transformed potential modulo generalized K\"ahler gauge transformations\footnote{Transformations of $K$ that do not affect the field eauations.}, and added the $\alpha$ term which represents a constant $B$ field, and hence will not change the geometry (which is torsionful). The corresponding action has the shift symmetry
\ber
\phi\to\phi+\lambda,~\chi \to \chi +\lambda~,
\eer
which we gauge using the Large Vector Multiplet (LVM) \cite{Lindstrom:2007sq}. This means that we couple the model to the three vector fields  $V^\phi, V^\chi$ and $V'$ and then introduce Lagrange multipliers $X_\phi, X_\chi$ and $X'$ that are linear combinations of semichiral fields and set the fieldstrengths of the LVM to zero. They read
\ber\nn\label{Xes}
&&X_\phi=\ihalf(\ell-\bar\ell-r+\bar r)\\[1mm]\nn
&&X_\chi=\ihalf(-\ell+\bar\ell-r+\bar r)\\[1mm]
&&X'=\half(\ell+\bar\ell-r-\bar r)~.
\eer 
It is convenient to redefine the LVM fields by a gague ransformation. Dualization is then the Legendre transformation
$(V^\phi, V^\chi, V')\to(X_\phi, X_\chi, X')$ obtained from
\ber\label{Vact}
-\half (V^\chi)^2-\alpha V^\chi V^\phi+\int^{V'}dq~{ln(1+e^q)} -V'X'-V^\phi X_\phi-V^\chi X_\chi~,
\eer
and results in the semichiral action (derived slightly differently in \cite{LRRUZ}, and discussed in \cite{Sevrin:2011mc}):
\ber\label{mod1}
-\frac 1{2\alpha^2}X_\phi^2+\frac 1{\alpha}X_\phi X_\chi -\int ^{X'}dq~ ln(e^q-1)~.
\eer
We expect the potential in \re{mod1} to correspond to a semichiral model with $(4,4)$ supersymmetry.

\subsection{Hyperk\"ahler duals}
\label{hkd}
Now, the semichiral dual of a (twisted) chiral model  with torsion and $(4,4)$ supersymmetry can be a (torsionless) hyperk\"ahler model when the target space is $4d$ if the original model has the LVM isometry in the coordinates where the Laplace equation takes the form \re{Lap}.  Namely, consider
\ber\nn\label{hkdual}
&&K=K\left(-\imath(\phi-\bar\phi),-\imath(\chi-\bar\chi), \phi+\bar\phi-\chi-\bar\chi\right)=:K(x,y,z)\\[1mm]
&& K_{\phi\bar\phi}+ K_{\chi\bar\chi}=0~\iff  K_{xx}+K_{yy}+2K_{zz}=0~.
\eer
As before, we may dualize to semichirals $(x,y,z)\to(u,v,w)$ using 
\ber
K(V^x,V^y,V^z)-uV^x-vV^y-wV^z~.
\eer
For ease of notation in what follows we use $(V^x,V^y,V^z)$ in place of $(V^\phi,V^\chi,V')$ and $(u,v,w)$  for $(X_\phi,X_\chi, X')$.
The dual (Legendre transformed) semichiral potential $\tilde K$ is found by integrating out the LVM $V$'s and solving them in terms of $(u,v,w)$:
\ber\nn
\tilde K(u,v,w)&=&K\left(V^x(u,v,w),V^y(u,v,w),V^z(u,v,w)\right)\\[1mm]
&&-uV^x(u,v,w)-vV^y(u,v,w)-wV^z(u,v,w)~.
\eer
Using the standard inverse relation that results from a Legendre transform we have   
\ber
K_{ab}= -(\tilde K_{ \hat b \hat a})^{-1}=: -\tilde K^{ \hat a\hat b}~,
\eer
where the unhatted indices represent the original coordinates and the hatted represent the dual.  Thus the condition for $(4,4)$ supersymmetry, the Laplace equation, in \re{hkdual} becomes 
\ber\label{lala}
K_{xx}+ K_{yy}+2K_{zz}=\tilde K^{uu}+\tilde K^{vv}+2\tilde K^{ww}=0~~.
\eer
This is in fact the form of hyperk\"ahler condition \re{Anticom} with $c=0$ takes in the $(u,v,w)$ coordinates.
In the original left and right coordinates, the only nontrivial condition that results from  \re{Anticom}  in $4d$ is
\ber
 (1+c)|\tilde K_{\ell r}|^2+(1-c)|\tilde K_{\ell \bar r}|^2=2\tilde K_{\ell \bar \ell}\tilde K_{r\bar r}~.
\eer
In $(u,v,w)$  coordinates this reads 
\ber\nn\label{HK}
&&2\tilde K_{uv}^2+\tilde K_{uw}^2+\tilde K_{vw}^2+c(\tilde K^2_{uw}-\tilde K^2_{vw}-\tilde K_{uu}\tilde K_{ww}+\tilde K_{vv}\tilde K_{ww})\\[1mm]
&&=2\tilde K_{uu}\tilde K_{vv}+\tilde K_{uu}\tilde K_{ww}+\tilde K_{vv}\tilde K_{ww}~.
\eer
It is a remarkable fact that {\em this is equivalent to}\footnote{ In coordinates 
$s:= (u+v)/2$ and $t= -(u-v)/2$ the condition becomes even simpler: $\tilde K^{ss}+\tilde K^{tt}+\tilde K^{ww}\propto c\tilde K^{st}$.}
\ber\label{remark}
\tilde K^{uu}+\tilde K^{vv}+2\tilde K^{ww}=4 c\left(\tilde K^{uu}-\tilde K^{vv}\right)~,
\eer
 so \re{lala} is satisfied (for $c=0$).
The relation \re{remark} may be verified using the following formula for the inverse of a $3\times3$ matrix:
\ber\nn
&&\left(\begin {array}{ccc}\tilde K_{uu}&\tilde K_{uv}&\tilde K_{uw}\\
\tilde K_{vu}&\tilde K_{vv}&\tilde K_{vw}\\
\tilde K_{wu}&\tilde K_{wv}&\tilde K_{ww}\end{array}\right)^{-1}\\[1mm]
&&=\frac 1 \Delta \left(\begin {array}{ccc}\tilde K_{vv}\tilde K_{ww}-\tilde K_{vw}^2&\tilde K_{vw}\tilde K_{uw}-\tilde K_{vu}\tilde K_{ww}&\tilde K_{vu}\tilde K_{vw}-\tilde K_{vv}\tilde K_{uw}\\
\tilde K_{vw}\tilde K_{uw}-\tilde K_{uv}\tilde K_{ww}&\tilde K_{uu}\tilde K_{ww}-\tilde K_{uw}^2&\tilde K_{uv}\tilde K_{uw}-\tilde K_{uu}\tilde K_{vw}\\
\tilde K_{vu}\tilde K_{wv}-\tilde K_{vv}\tilde K_{uw}&\tilde K_{uv}\tilde K_{uw}-\tilde K_{uu}\tilde K_{vw}&\tilde K_{uu}\tilde K_{vv}-\tilde K_{uv}^2\end{array}\right)~,
\eer
where $\Delta$ denotes the determinant of the original matrix. 

The conclusion of this subsection is thus that if a (twisted) chiral $4d$ sigma model satisfied the Laplace equation {\em and} has LVM (shift) isometry  in these coordinates, it has semichiral dual which is hyperk\"ahler.

This explains the geometry underlying the hyperk\"ahler constructions of \cite{Bogaerts:1999jc}.
\section{Extended supersymmetry in $(1,1)$ superspace}

The conclusions in subsec.\ref{hkd} do not apply to our model in \re{mod1}. In the coordinates where the Laplace equation takes the form \re{Lap}, the LVM isometry is not realized as a translation symmetry, and conversely, in the coordinates where the LVM isometry is realized as translations the condition for extended supersymmetry takes the form \re{Lap2}.
Indeed this model has torsion, as can be directly checked and it has extended supersymmetry inherited from the potential \re{Bilp2}. 

To address the question of how the additional supersymmetry is realized it is useful to descend to $(1,1)$ superspace, in light of the negative results in section.\ref{Three}.
\subsection{$(1,1)$ keeping the auxiliary spinors }
The potential \re{oneact} results in the following $(1,1)$ sigma model Lagrangian
\ber\label{completed}
{\cal L} = D_+X^ i{E}_{ij}(X)D_-X^j+\Psi_+^RK_{RL}\Psi_-^L := {\cal L}_1+{\cal L}_2~,
\eer
where the $\Psi$s are related to the components in \re{onecomps} as
\ber\nn\label{bigpsi}
&&\Psi_+^r:=\psi_+-D_+X^AJ_{(+)A}^r\\[1mm]
&&\Psi_-^\ell:=\psi_--J_{(-)A}^\ell D_-X^A~,
\eer
and their complex conjugate ($R=:(r,\bar r)~, L=:(\ell,\bar\ell) \Rightarrow i=(L,R)$). Integrating out the auxiliary spinors sets $\Psi=0$, but here we keep them and investigate thir role in  extended supersymmetry transformations 
forming an $SU(2)$ algebra\footnote{Corresponding to $(+)$-supersymmetries. The general case also involves 
$(-)$-supersymmetries.}. Explicitly, the extended supersymmetries are generated by complex structures \cite{Gates:1984nk}
\ber\label{su2}
I_{(+)}^{(\textfrak a)}I_{(+)}^{(\textfrak b)}=-\delta^{\textfrak a\textfrak b}+\epsilon^{{\textfrak a \textfrak b \textfrak c}}I_{(+)}^{({\textfrak c})}
\eer
with $\textfrak a=1,2,3$ and the identification $I_{(+)}^{(3)}:=J_{(+)}$.  Starting from one of the complex structures $I^{(\frak a)}=: I^i_{~j}$ and assuming that the corresponding transformation of $X^ i$ yields a symmetry of ${\cal L}_1$, we ask under what condition the transformation can be extended to include $\Psi$ compatible with \re{bigpsi} and leaving the full Lagrangian invariant (up to total derivatives). At the $(2,2)$ level, such a transformation would correspond to a transformation of the semichiral fields either leading to hyperk\"ahler geometry or  somehow missed in our $(2,2)$ analysis.

A general ansatz for the non manifest $(+)$-supersymmetry is (dropping the $(+)$ index on $I$):
\ber\nn\label{exinv}
&&\delta X^i=\epsilon^+\left[I^i_{~j}D_+X^j+M^i_{~R}\Psi^R_+\right]~,
\eer
where $M$ is a $4\times 4$ matrix with $M^i_{~L}=0$. The $\Psi$-transformations follows from transforming  both sides in \re{bigpsi} and read 
\ber\nn\label{psinv}
&&\delta\Psi^{\dot r}_+=\epsilon^+\left[\left(I^{\dot r}_{~r}-[M,J_{(+)}]^{\dot r}_{~r}\right)D_+\Psi^r_+
+\left(I^{\dot r}_{~j,r}+{\cal M}(M,J_{(+)})^{\dot r}_{rj}\right)D_+X^j\Psi^r_+\right.\\[1mm]\nn
&&~~~~~~~~~~~~~~~\left. -M^{\dot r}_{~r,r'}\Psi^{r'}_+\Psi^r_+\right]\\[1mm]\nn
&&\delta\Psi^{\dot \ell}_-=\epsilon^+\left[-[M,J_{(-)}]^{\dot \ell}_{~r}D_-\Psi^r_++{\cal M}(M,J_{(-)})^{\dot \ell}_{rj}D_-X^j\Psi^r_+\right.\\[1mm]\nn
&& ~~~~~~~~~~~~~+(I^{\dot \ell}_{~\ell}-M^{\dot \ell}_{~r}J^r_{(+)\ell})D_+\Psi^\ell_-+(I^{\dot \ell}_{~j,\ell}-M^{\dot \ell}_{~r}J^r_{(+)j,\ell})D_+X^j\Psi^\ell_-\\[1mm]\nn
&&~~~~~~~~~~~~~~\left.-M^{\dot \ell}_{~r,\ell}\Psi^{\ell}_-\Psi^r_++\left([I,J_{(-)}]^{\dot \ell}_{~j}-M^{\dot \ell}_{~r}[J_{(+)},J_{(-)}]^r_{~j}\right)\nabla^{(-)}_+D_-X^j\right]~,\\[1mm]
&&
\eer
and their complex conjugate expressions. Here overdot denotes the free index, covariant derivatives are defined in \re{covd} and
the Magri-Morosi concomitant for two endomorphisms $I$ and $J$ reads \cite{MM}
\ber
{\cal M}(I,J)^{A}_{BD}=I^F_{~B}J^A_{~D,F}-J^F_{~D}I^A_{~B,F}-I^A_{~F}J^F_{~D,B}+J^A_{~F}I^F_{~B,D}~.
\eer
The algebra of transformations \re{exinv} and \re{psinv} close on-shell where $\Psi=0$, $\nabla^{(-)}_+D_-X^j=0$ and the $X$ transformations close. The issue is invariance of the action. 

Invariance of the action requires $M$ to satisfy a number of conditions which we now list:
First, raising and lowering indices on $M$ with  $K_{RL}$\footnote{This is just a notational convenience and does not necessarily identify $K_{RL}$ as a metric}, 
\ber\nn\label{morecalc}
&&{M_{L[R,\dot R]}}{-}{M_{[R\dot R],L}=0}\\[1mm]\nn
 &&{M_{[R\dot R]}={{-}}\half K_{\dot R \dot L}[I_{(+)},J_{(-)}]^{ \dot L}_{~j}{g}^{jL}}K_{LR}\\[1mm]\nn
 &&{M^R_{~\dot R}={{-}}\half K_{\dot RL}[I_{(+)},J_{(-)}]^{ L}_{~j}g^{jR}}\\[1mm]\nn
&&{-}{K_{(\dot R |L|}[M,J_{(-)}]^L_{~R)}=[J,M]_{(\dot R R)}+C_{(\dot R |R|}M^R_{~R)}=0}\\[1mm]\nn
	&&{K_{[\dot R |L|}{\cal M}(M,J_{(-)})^{ L}_{R]j}D_-X^j=-\half D_-(K_{[\dot R |L|}[M,J_{(-)}]^L_{~R]})}\\[1mm]\nn
&&{K_{LR}I^R_{~\dot R}-K_{\dot R L}I^L_{~L}+K_{LR}[\tilde J,M]^R_{~\dot R}+C_{LL}K^{LR}M_{[R\dot R]}=0}~,\\[1mm]\nn
&&{{\left(I^R_{~j,\dot R}+{\cal M}(M,J_{(+)})^R_{\dot Rj}\right)K_{RL}+K_{\dot R L k}I^k_{~j}+K_{\dot R \dot L}\left(I^{\dot L}_{~j,L}-M^{\dot L}_{~R}J^R_{(+)j,L}\right)}}\\[1mm]
&&{=\left((I^{\dot L}_{~L}-M^{\dot L}_{~R}J_{(+)L}^R)K_{\dot L\dot R}\right)_{,j}}~,
\eer
where $(\tilde J)^R_{~\dot R}:=K^{RR'}JK_{R'\dot R}$ an we have used \re{morecalc} and the explicit form of $J_{(+)}$ from \cite{Lindstrom:2005zr} in the last equation, as well as the definitions
\ber
C_{LL}:=[J,K]_{LL}=2\imath\left(
\begin{array}{cc}0&K_{\ell,\bar \ell}\\
-K_{\bar \ell\ell}&0
\end{array}\right)=-2 K_{\ell,\bar \ell}\sigma_2~,~~~~C_{RR}=-2 K_{r,\bar r}\sigma_2~,
\eer
with the relations involving the Pauli matrix being particular to $4d$ target space.

The set of conditions \re{morecalc} looks forbidding. Nevertheless we know that it has to be satisfied by $I=J_{(+)}$, corresponding to the second supersymmetry which is manifest in $(2,2)$ superspace. Indeed it is, as seen below.

The conditions are also expected to be satisfied for some hyperk\"ahler manifolds, as discussed in sec.\ref{Three}. Testing this on the hyperk\"ahler
complex structures in \re{hks}, we find that the relations \re{morecalc} determine $M$ in the three cases according to
\ber\nn\label{HKE}
J^{(3)}: && M^R_{~\dot R} = \delta^R_{~\dot R}~,~~M_{[\dot R R]}=0\\[1mm]\nn
J^{(1)}: && M^R_{~\dot R} = \frac {c~\delta^R_{~\dot R}}{\sqrt {1-c^2}}~,~~\\[1mm]\nn
J^{(2)}: && M^R_{~\dot R}K_{L R} = -\frac 1 {\sqrt{1-c^2}} K_{\dot R L}J_{(-)L}^L=-\frac 1 {\sqrt{1-c^2}}JK_{\dot R L}\\[1mm]\nn
&& M_{[\dot R R]} =-\frac 1 {\sqrt{1-c^2}} K_{\dot R L}J_{(-)R}^L= -\frac 1 {\sqrt{1-c^2}} C_{\dot R R}\\[1mm]
\eer
Each case satisfies the relation in \re{morecalc} (provided that $c$ is constant), again using the relations for $J_{(\pm)}$ from \cite{Lindstrom:2005zr} . We now want to test the relations on our semichiral version of $SU(2)\otimes U(1)$.
\subsection{The additional complex structures for $SU(2)\otimes U(1)$}

The additional supersymmetries of \re{Bilp1} may be found in \cite{Gates:1984nk}. In the new coordinates where the potential is \re{Bilp2} they read 
\ber\nn\label{2ndsusy}
&&\delta \phi=e^{\bar\chi-\phi}\bar\epsilon^+\bar{\mathbb{D}}_+{\bar\chi}+e^{\chi-\phi}\bar\epsilon^-\bar{\mathbb{D}}_-\chi\\[1mm]\nn
&&\delta {\bar\phi}=e^{\chi-\bar\phi}\epsilon^+{\mathbb{D}}_+{\chi}+e^{\bar\chi-\bar\phi}\epsilon^-\mathbb{D}_-{\bar\chi}\\[1mm]\nn
&&\delta {\chi}=-e^{\bar\phi-\chi}\bar\epsilon^+{\mathbb{D}}_+{\bar\phi}-e^{\phi-\chi}\epsilon^-{\mathbb{D}}_-{\phi}\\[1mm]
&&\delta {\bar\chi}=-e^{\phi-\bar\chi}\epsilon^+{\mathbb{D}}_+{\phi}-e^{\bar\phi-\bar\chi}\bar\epsilon^-\bar{\mathbb{D}}_-{\bar\phi}~.
\eer
These relations survive in the $(1,1)$ reduction with ${\mathbb{D}}_{\pm}\to D_\pm$.
From the general formula of the supersymmetry transformation of a field $\varphi$ and \re{2ndsusy} we  read off the additional complex structures:
\ber
\delta \varphi=\half\left[ \left (I^{(1)}_{(\pm)}+iI^{(2)}_{(\pm)}\right)\epsilon^\pm D_\pm\varphi+\left (I^{(1)}_{(\pm)}-iI^{(2)}_{(\pm)}\right)\bar\epsilon^\pm D_\pm\varphi\right]~.
\eer\nn
For the $I^{(\textfrak a)}_{(+)}$ we find (in a $(\phi,\chi,\bar\phi, \bar\chi)$  basis )
\ber
&&I^{(\textfrak a))}_{(+)}=\left(\begin{array}{cc}
0&\mathbb{A}^{(\textfrak a)}\\
-(\mathbb{A}^{(\textfrak a)})^{-1}&0\end{array}\right)
\eer
for $\textfrak a=1,2$, with
\ber\nn\label{Adef}
&&\mathbb{A}^{(1)}=\left(\begin{array}{cc}
0&e^{\bar\chi-\phi}\\
e^{\chi-\bar\phi}&0\end{array}\right)\\[2mm]
&&\mathbb{A}^{(2)}=\left(\begin{array}{cc}
0&ie^{\bar\chi-\phi}\\
-ie^{\chi-\bar\phi}&0\end{array}\right)~,
\eer
and with $I^{(3)}_{(+)}=J$, the canonical complex structure. To find these complex structures on the semichiral side, we note that unlike the general case for duality, at the $(1,1)$ level, there must be a coordinate transformation that take us from the (twisted) chiral coordinates to the semichiral coordinates, since both coordinatize the same geometric object. 

A $(1,1)$ superspace analysis of the duality \re{Bilp2} to \re{mod1} yields the transformation \cite{Lindstrom:2014bra}
\ber\nn\label{ctfs}
&&i(\bar \phi-\phi)=-i (r-\bar r)\\[1mm]\nn
&&i(\bar\chi-\chi)=\ihalf(\ell-\bar\ell-r+\bar r)\\[1mm]
&&\phi+\bar\phi-\chi-\bar\chi=ln(e^{\half(\ell+\bar \ell-r-\bar r)}-1)~.
\eer
This transformation is not complete, however. One more relation is needed.  As described in \cite{Lindstrom:2014bra}, it turns out to be most convenient to identify the full coordinate transformation not in $(L,R)$ semichiral coordinates, but in so called $(X,Y)$ coordinates where the $J_{(+)}$ derived from the semichiral model is canonical   \cite{Lindstrom:2005zr} \cite{Bogaerts:1999jc}. We {\em require that it stays invariant} under the coordinate transformation and thus is mapped to the $J_{(+)}$ in the (twisted) chiral model. As discussed in  \cite{Lindstrom:2014bra}, this condition is natural but not necessarily the only possibility. In \cite{Ivanov:1994ec} the transformations of complex structures under T-duality is described. The transformed complex structures are obtained by finding the supersymmetry for the vector fields in the first order action for the $(1,1)$ dualization. Using this method directly is difficult, since the relation to the $(2,2)$ coordinate fields is needed as additional data.  It should also be noted that the derivation is not directly applicable in the present case where the first order action contains more vector fields and auxiliaries. A modified derivation can presumably be worked out using the T-duality relations in \cite{Lindstrom:2007sq}. Here it is sufficient to note that the present route yields the extended supersymmetry of the semichiral model corresponding to that of the (twisted) chiral model.

For $\textfrak a =1,2$, the expression for $I^{(\textfrak a)}_{(+)}$,  in $(L,R)$ coordinates turns out to be\footnote{The full coordinate transformation is given in \cite{Lindstrom:2007sq}.}
\ber\nn
I^{(\textfrak a)}_{(+)}=
&&\frac 1 {4N}\left\{\left(\begin{array}{cc}E&-M\\
e^{-X}&-e^{-X}E
\end{array}\right)\otimes {\mathbb{A}}^{(\textfrak a)}+\left(\begin{array}{cc}M&-ME\\
e^{-X}E&-e^{-X}E^2
\end{array}\right)\otimes {\mathbb{A}}^{(\textfrak a)}\sigma_1\right.\\[2mm]
&&~~~~~~\left. +\frac N{e^X}\left[
\left(\begin{array}{cc}0&0\\
1&-E
\end{array}\right)\otimes \bar{\mathbb{A}}^{(\textfrak a)}+\left(\begin{array}{cc}0&0\\
E&-1
\end{array}\right)\otimes \bar{\mathbb{A}}^{(\textfrak a)}\sigma_1
\right]\right\}~,
\eer
where 
\ber
 -\frac 1 {4N}:=K_{\ell\bar\ell}~,~~~-\frac E {4N}:=K_{\ell\ell}~,~~~\frac M {4N}:=K_{r\bar r}~,
\eer
$K$ is given in \re{mod1} and $X$ is defined in \re{Xes} (dropping the prime). 

As explained in  \cite{Lindstrom:2007sq}, these $I^{(\textfrak a)}_{(+)}$ give matrices $M^i_{~J}$ that do not satisfy the conditions \re{morecalc} and hence the corresponding transformations cannot be lifted to transformations on semichiral fields.

\section{Conclusions}

Using a novel $(1,1)$ formulation of semichiral sigma models where the auxiliary spinors are kept, conditions for extra supersymmetries that can be lifted to a $(2,2)$ semichiral formulation are derived. For a  $4d$ target space these conditions are shown to be met in supersymmetries generated by a hyperk\"ahler set of complex structures including the second supersymmetry generated by $J_{(+)}$. This agrees with the $(2,2)$ result that hyperk\"ahler is an allowed targetspace geometry when extra supersymmetries are realized as transformations of the semichiral fields \cite{Goteman:2009ye}. In the paper \cite{Goteman:2009ye} it is also argued that the $SU(2)\otimes U(1)$ WZW model shows that hyperk\"ahler is not exhaustive, since there is a (twisted) chiral formulation with manifest $(4,4)$ which has a semichiral dual.

To investigate the counterexample, we derive the extra right complex structures and map them to the semichiral coordinates in the $(1,1)$ formulation, thus displaying the extra right supersymmetries for the semichiral model for the first time (for the $(1,1)$ sigma model with auxiliary spinors integrated out).  We then derive the transformations of the auxiliary spinors for this case and check if the symmetry can be lifted to transformations on semichirals in $(2,2)$ superspace. The result is negative even for this case of extra right transformations only.

The resolution of the counterexample conundrum is thus that {\em the extra supersymmetries are there in the $(1,1)$ formulation of the semichiral model with zero auxiliary spinors, but are incompatible with nonzero auxiliaries}. Another way of saying this at the $(2,2)$ level is that the chirality conditions on the semichiral fields is incompatible with a realization of the extra supersymmetries as transformations of those fields. Presumably additional $(2,2)$ auxiliary fields are needed.
\bigskip

{\bf Acknowledgement}: Discussions with M. Ro\v cek at various stages of this work are gratefully acknowledged.
Supported in part by VR grant 621-2013-4245. I am grateful to the organizers of the Corfu Summer Institute 2014 for the invitation.\\

\end{document}